\documentclass[dvips]{acta}
\usepackage{supertabular,lscape,epsfig}
\usepackage{amssymb}
\usepackage{amsmath}
\DeclareSymbolFont{ppa}{OT1}{ppl}{m}{it}
\DeclareMathSymbol{\vv}{\mathalpha}{ppa}{'166}

\newfont{\hb}{rphvb at 10pt}
\newfont{\hbo}{rphvbo at 10pt}
\newfont{\bitt}{rptmbi at 12pt}
\newfont{\bits}{rptmbi at 11pt}

\SetPages{1}{19}

\SetVol{63}{2013}

\begin{document}

\newcommand{\TabApp}[2]{\begin{center}\parbox[t]{#1}{\centerline{
  {\bf Appendix}}
  \vskip2mm
  \centerline{\small {\spaceskip 2pt plus 1pt minus 1pt T a b l e}
  \refstepcounter{table}\thetable}
  \vskip2mm
  \centerline{\footnotesize #2}}
  \vskip3mm
\end{center}}

\newcommand{\TabCapp}[2]{\begin{center}\parbox[t]{#1}{\centerline{
  \small {\spaceskip 2pt plus 1pt minus 1pt T a b l e}
  \refstepcounter{table}\thetable}
  \vskip2mm
  \centerline{\footnotesize #2}}
  \vskip3mm
\end{center}}

\newcommand{\TTabCap}[3]{\begin{center}\parbox[t]{#1}{\centerline{
  \small {\spaceskip 2pt plus 1pt minus 1pt T a b l e}
  \refstepcounter{table}\thetable}
  \vskip2mm
  \centerline{\footnotesize #2}
  \centerline{\footnotesize #3}}
  \vskip1mm
\end{center}}

\newcommand{\MakeTableApp}[4]{\begin{table}[p]\TabApp{#2}{#3}
  \begin{center} \TableFont \begin{tabular}{#1} #4 
  \end{tabular}\end{center}\end{table}}

\newcommand{\MakeTableSepp}[4]{\begin{table}[p]\TabCapp{#2}{#3}
  \begin{center} \TableFont \begin{tabular}{#1} #4 
  \end{tabular}\end{center}\end{table}}

\newcommand{\MakeTableee}[4]{\begin{table}[htb]\TabCapp{#2}{#3}
  \begin{center} \TableFont \begin{tabular}{#1} #4
  \end{tabular}\end{center}\end{table}}

\newcommand{\MakeTablee}[5]{\begin{table}[htb]\TTabCap{#2}{#3}{#4}
  \begin{center} \TableFont \begin{tabular}{#1} #5 
  \end{tabular}\end{center}\end{table}}

\newfont{\bb}{ptmbi8t at 12pt}
\newfont{\bbb}{cmbxti10}
\newfont{\bbbb}{cmbxti10 at 9pt}
\newcommand{\uprule}{\rule{0pt}{2.5ex}}
\newcommand{\douprule}{\rule[-2ex]{0pt}{4.5ex}}
\newcommand{\dorule}{\rule[-2ex]{0pt}{2ex}}
\def\thefootnote{\fnsymbol{footnote}}
\begin{Titlepage}
\vglue-9pt
\Title{Supernovae and Other Transients in the OGLE-IV \\ Magellanic Bridge Data\footnote{Based on
observations obtained with the 1.3-m Warsaw telescope at the Las Campanas
Observatory of the Carnegie Institution for Science.}}

\Author{S.~~K~o~z~³~o~w~s~k~i$^{1}$,~~~
A.~~U~d~a~l~s~k~i$^1$,~~~
£.~~W~y~r~z~y~k~o~w~s~k~i$^{1,2}$,~~~
R.~~P~o~l~e~s~k~i$^{1,3}$,
P.~~P~i~e~t~r~u~k~o~w~i~c~z$^1$,~~~
J.~~S~k~o~w~r~o~n$^1$,~~~
M.\,K.~~S~z~y~m~a~ñ~s~k~i$^1$,~~~
M.~~K~u~b~i~a~k$^1$,
G.~~P~i~e~t~r~z~y~ñ~s~k~i$^{1,4}$,~~
I.~~S~o~s~z~y~ñ~s~k~i$^1$~~~
and~~~
K.~~U~l~a~c~z~y~k$^1$
}
{
$^1$ Warsaw University Observatory, Al.~Ujazdowskie~4, 00-478~Warszawa, Poland\\
e-mail:(simkoz,udalski,lw,rpoleski,pietruk,jskowron,msz,mk,pietrzyn,soszynsk,kulaczyk)
@astrouw.edu.pl\\
$^2$ Institute of Astronomy, University of Cambridge, Madingley Road, Cambridge CB3 0HA, UK\\ 
$^3$ The Ohio State University, 140 W. 18th Avenue, Columbus, OH 43210, USA\\
$^4$ Departamento de Astronomia, Universidad de Concepción, Casilla 160-C, Concepción, Chile}
\vspace*{-5pt}
\Received{January 17, 2013}
\end{Titlepage}

\vspace*{-7pt}
\Abstract{We analyze two years (mid-2010 to mid-2012) of OGLE-IV data 
covering $\approx65$~deg$^2$ of the Magellanic Bridge (the area
between the Magellanic Clouds) and find 130 transient events including
126 supernovae (SNe), two foreground dwarf novae and another two
SNe-like transients that turned out to be active galactic nuclei
(AGNs). We show our SNe detection efficiency as a function of SN peak
magnitude based on available SNe rate estimates. It is 100\% for SNe
peak magnitudes $I<18.8$~mag and drops to 50\% at
$I\approx19.7$~mag. With our current observing area between and around
the Magellanic Clouds ($\approx600$~deg$^2$), we expect to find 24 SNe
peaking above $I<18$~mag, 100 above $I<19$~mag, and 340 above
$I<20$~mag, annually.  We briefly introduce our on-line near-real-time
detection system for SNe and other transients, the OGLE Transient
Detection System.}{supernovae: general -- Magellanic Clouds -- Surveys}

\vspace*{-9pt}
\Section{Introduction}
\vspace*{-5pt}
Supernovae (SNe) are gigantic stellar explosions marking the ultimate
death of massive stars.  They are extremely luminous, often outshining
all stars of the host galaxy for a short period of time, making them
excellent probes of the distant Universe, to $z\approx1.7$ for SNe
Type~Ia (Riess \etal 2001), $z\approx2.4$ for SNe Type~II (Cooke \etal
2009), and $z\approx3.9$ for super-luminous SNe (Cooke \etal 2012).
Type~Ia SNe, exploding white dwarfs, can be used as ``standardizable''
candles (\eg Phillips 1993). This yardstick redefined our view of the
Universe, playing a major role in discovering the accelerating
expansion of the Universe (\eg Riess \etal 1998, Perlmutter \etal
1999). Recent studies use total samples of $\approx600$ SNe Type~Ia to
constrain evolution of the Universe (\eg Conley \etal 2011,
Suzuki \etal 2012). Our understanding of the progenitors of SNe
Type~Ia, their hosts, and the connection between them remains limited
as illustrated by the continuing debate over the relative roles of the
single-degenerate (SD) and double-degenerate (DD) channels for
producing them.

While core collapse SNe (ccSNe -- Type II and Ibc) may also be
``standardizable'' candles (\eg Hamuy and Pinto 2002, Poznanski, Nugent and
Filippenko 2010), they are primarily studied as probes of the evolution and
death of massive stars (\eg Woosley and Weaver 1986, Smartt 2009) and for
their role in providing the ``feedback'' that regulates star formation (\eg
Silk 2005). Some important open questions are the death of higher mass
progenitors of ccSNe (Kochanek \etal 2008, Smartt \etal 2009, Jennings
\etal 2012), correlations of properties with metallicity (Prieto, Stanek,
and Beacom 2008, Koz³owski \etal 2010b, Stoll \etal 2011), apparent
correlations of pre-SN outbursts with the SN (\eg Pastorello \etal 2007)
and the nature of the so-called SN impostors (Smith \etal 2011, Kochanek,
Szczygie³ and Stanek 2012). Many of these issues can be addressed by
unbiased surveys for ccSNe.

In the early days of the Optical Gravitational Lensing Experiment
(OGLE), we also searched for SNe in parallel with our successful
microlensing searches (Udalski \etal 1993). Due to the small area of
the CCD camera and the limited number of observing nights awarded on
the 1.0~m Swope telescope this was not a competitive survey. We
returned to the problem while implementing ``The New Objects in the
OGLE-III Sky'' real-time system (NOOS; Udalski 2003) as part of the
larger area OGLE-III survey, using the 1.3 m Warsaw telescope,
discovering several dozen SNe over the years 2003--2005 (\eg Udalski
2004a,b).

In its fourth phase, the OGLE survey (OGLE-IV), also conducted on the
1.3~m Warsaw telescope, uses a 32 CCD mosaic camera covering
1.4~deg$^2$. One of the OGLE-IV survey goals is to find variable stars
in a wide area ($\approx600$~deg$^2$) around and between the
Magellanic Clouds in order to study the spatial distribution of stars
including Cepheids and RR~Lyr stars (Soszyñski \etal 2013, in
preparation). On average, these areas have low stellar density and
many galaxies are readily detected in our images (see
Soszyñski \etal 2012). With an observing cadence of 2--3 days, we
find not only ``regular'' variable stars but also all types of
transient events and active galactic nuclei (AGNs). While the OGLE-IV
survey of the Magellanic Clouds was not specifically designed to find
SNe, we are finding large numbers of them as a by-product (\eg
Wyrzykowski, Udalski and Koz³owski 2012). Our goal is not to
directly compete with dedicated SNe surveys, such as the Catalina
Real-Time Transient Survey (CRTS; Drake \etal 2009), the Palomar
Transient Factory (Law \etal 2009), the Supernova Legacy Survey
(Astier \etal 2006), the Lick Observatory Supernova Search (LOSS,
Li \etal 2011a,b), the Hubble Space Telescope Cluster Supernova Survey
(Suzuki \etal 2012), the CANDELS survey (Grogin \etal 2011,
Koekemoer \etal 2011), but simply to contribute to the SN field with
discoveries.

In this paper, we are interested in finding SNe amongst the variable
OGLE-IV objects in the vicinity of the Magellanic Clouds. In Section~2, we
describe the collected data, their analysis, and the methods to find
SNe. The expected SNe numbers from known SNe rates as well as our detection
efficiency is presented in Section~3. In Section~4, we describe the method
of discriminating SNe from AGNs. A brief introduction to our near-real-time
transient detection system can be found in Section~5. The paper is
summarized and the data availability is described in Section~6.

\vspace*{-7pt}
\Section{Observational Data}
\vspace*{-4pt}
The OGLE-IV survey is conducted with the 1.3~m Warsaw telescope
located at the Las Campanas Observatory in Chile, operated by the
Carnegie Institution for Science. In the current, fourth phase of the
OGLE survey, the telescope is equipped with a mosaic camera composed
of 32 CCD detectors, each having $2048\times4102$ pixels, totaling 268
mega-pixels. The new camera covers approximately 1.4~deg$^2$ of the
sky with a scale of 0.26~arcsec/pixel. In this paper, we analyze the
\begin{figure}[htb]
\centerline{\includegraphics[width=12cm]{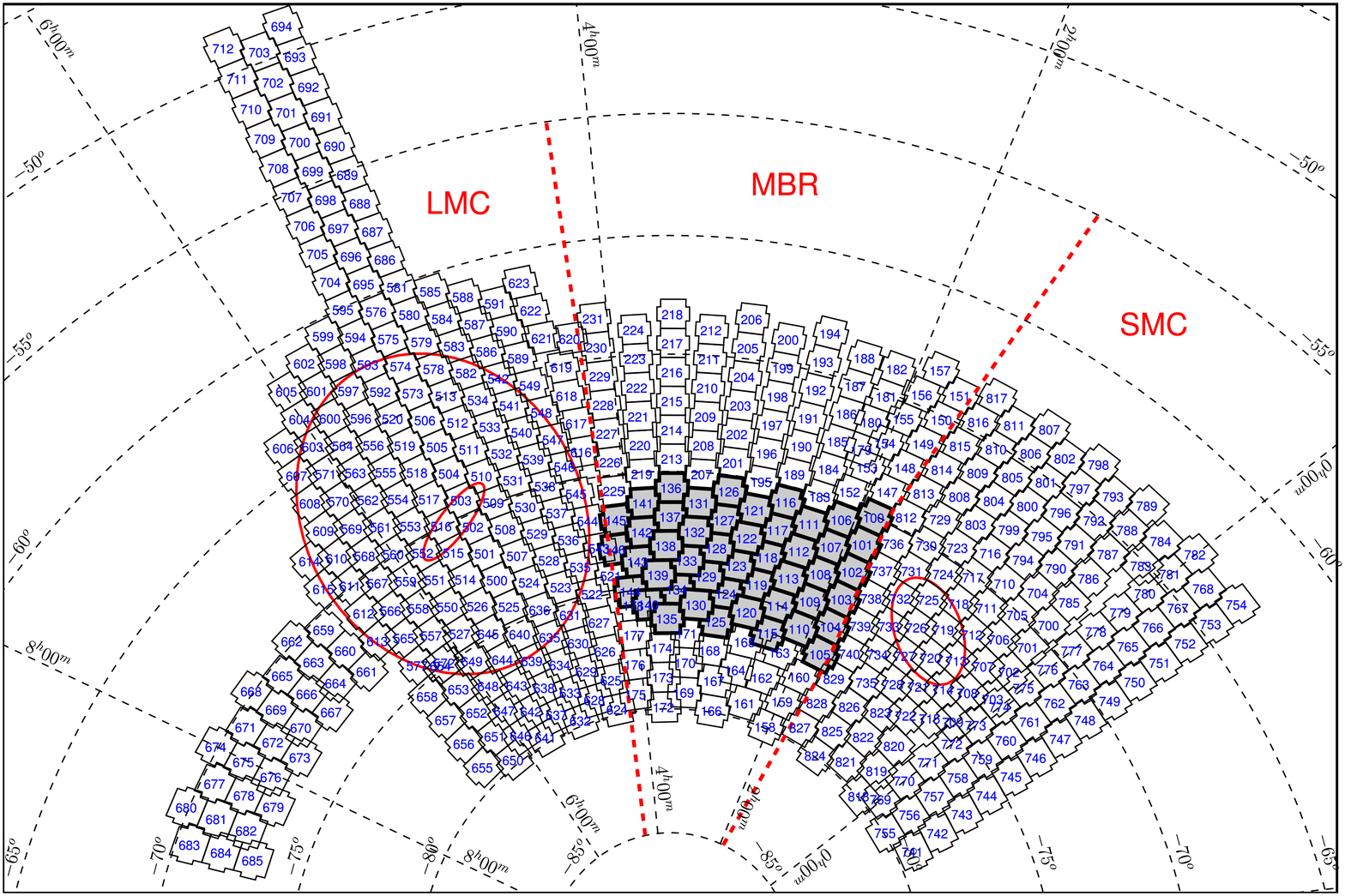}}
\vspace*{2mm}
\FigCap{Area of the Magellanic Clouds that has been monitored by 
OGLE-IV since 2012 is shown. Each labeled square-like shape
corresponds to one 1.4~deg$^2$ OGLE-IV field. The early (2010--2012)
monitored area of the Magellanic Bridge, analysis of which we report
in this paper, is marked with thick lines ($\approx65$~deg$^2$). The
total current observed area is $\approx600$~deg$^2$. We schematically
show the SMC (ellipse on the right) and the LMC's size and its bar
(large and small ellipses on the left, respectively). Two nearly
vertical dashed lines divide the LMC ({\it left}), MBR ({\it middle}),
and SMC ({\it right}) areas.}
\end{figure}
first two years (mid-2010 to mid-2012) of the Magellanic Bridge
OGLE-IV data. This region consists of 47 slightly overlapping fields,
covering a total of $\approx65$~deg$^2$ (Fig.~1). The median number of
{\it I}-band points per $\approx$8-month-long season is 91, with an
average cadence of 2.7~days. The seasonal gaps last $\approx4$ months.

The CCD frames are reduced on-the-fly at the telescope with the
best-available bias and flat-field images. They are then fed to our own
automatic difference image analysis (DIA technique; Wo¼niak 2000)
photometric pipeline (Udalski \etal 2008). Each image of a given field is
aligned to the corresponding template (reference) image, an average of high
quality images, and the template is convolved and scaled to match the point
spread function (PSF) flux and background, and then subtracted. The only
remaining objects on the resulting difference image are real transient and
variable objects, asteroids, satellite trails, cosmic rays, and noise.

\begin{figure}[htb]
\centerline{\includegraphics[width=8.3cm]{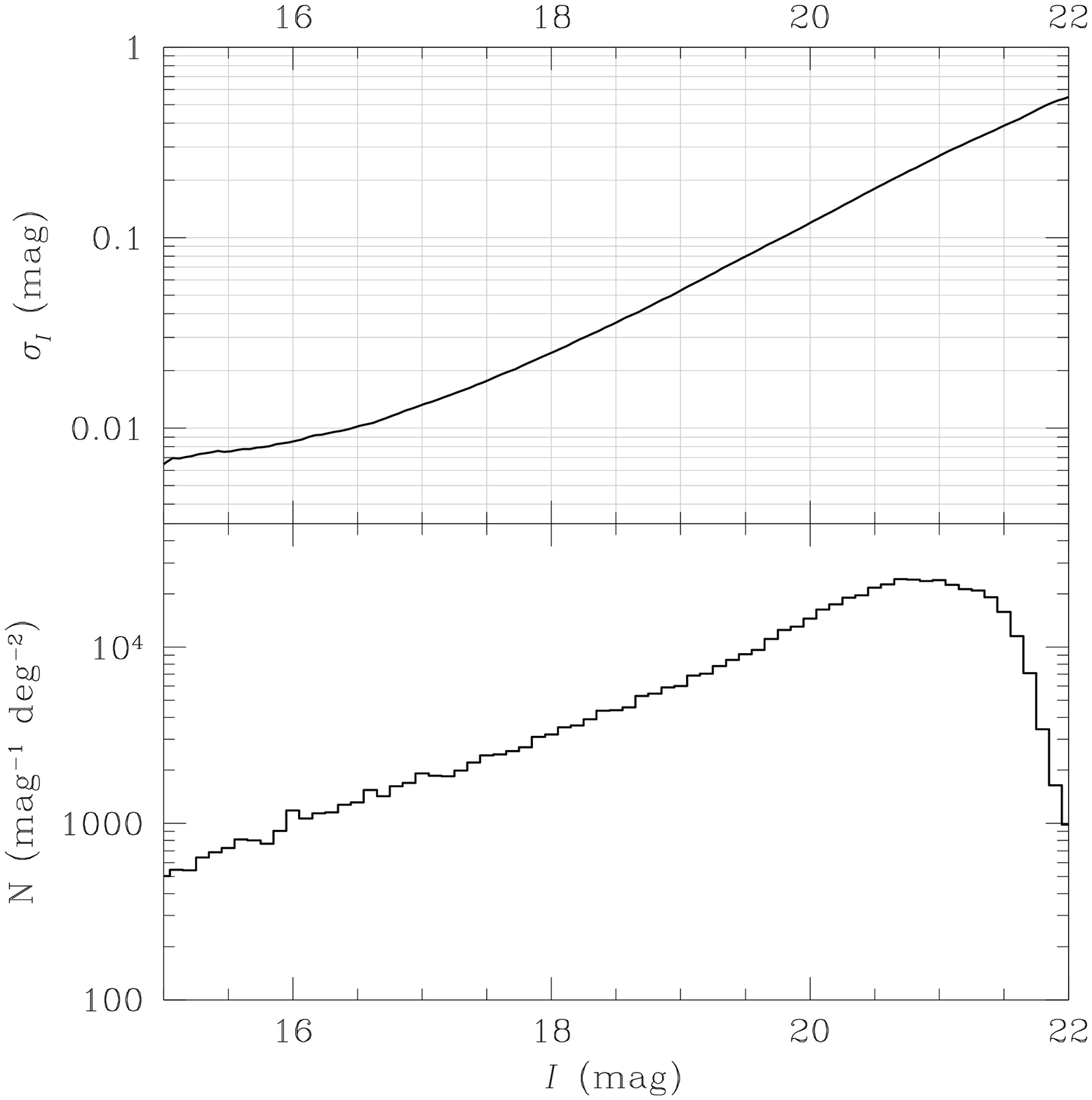}}
\vspace*{3mm}
\FigCap{Data statistics for central OGLE-IV MBR fields (MBR118 and MBR123). 
{\it Top:} Smoothed (three 0.05 mag bins) median dispersion for
non-variable light curves as a function of {\it I}-band magnitude.
{\it Bottom:} Histogram of the number of detected objects per
magnitude per deg$^2$. The detection of objects on the template image
is complete to $I\approx20.7$~mag. It then flattens out to
$I\approx21.3$~mag, and drops quickly to zero at about
$I\approx22.0$~mag. Transients from the database of new objects
(appearing from below background) therefore come from
progenitors/galaxies fainter than this limit.}
\end{figure}
The OGLE pipeline builds two databases. The first one is directly
dependent on the template image. We detect all objects on the template
image and then create light curves for each source by adding the flux
measured for it on the difference image to the template flux. This is
our standard database. The second database consists of new objects
that are detected only on the difference images and have no
counterparts on the template image within 0\zdot\arcs5. In Fig.~2, we
present the OGLE-IV data quality -- the median magnitude dispersion
for non-variable objects as a function of magnitude and the number of
detected objects on template images as a function of magnitude in low
stellar density Magellanic Bridge fields.

\begin{figure}[htb]
\centerline{\includegraphics[width=11cm]{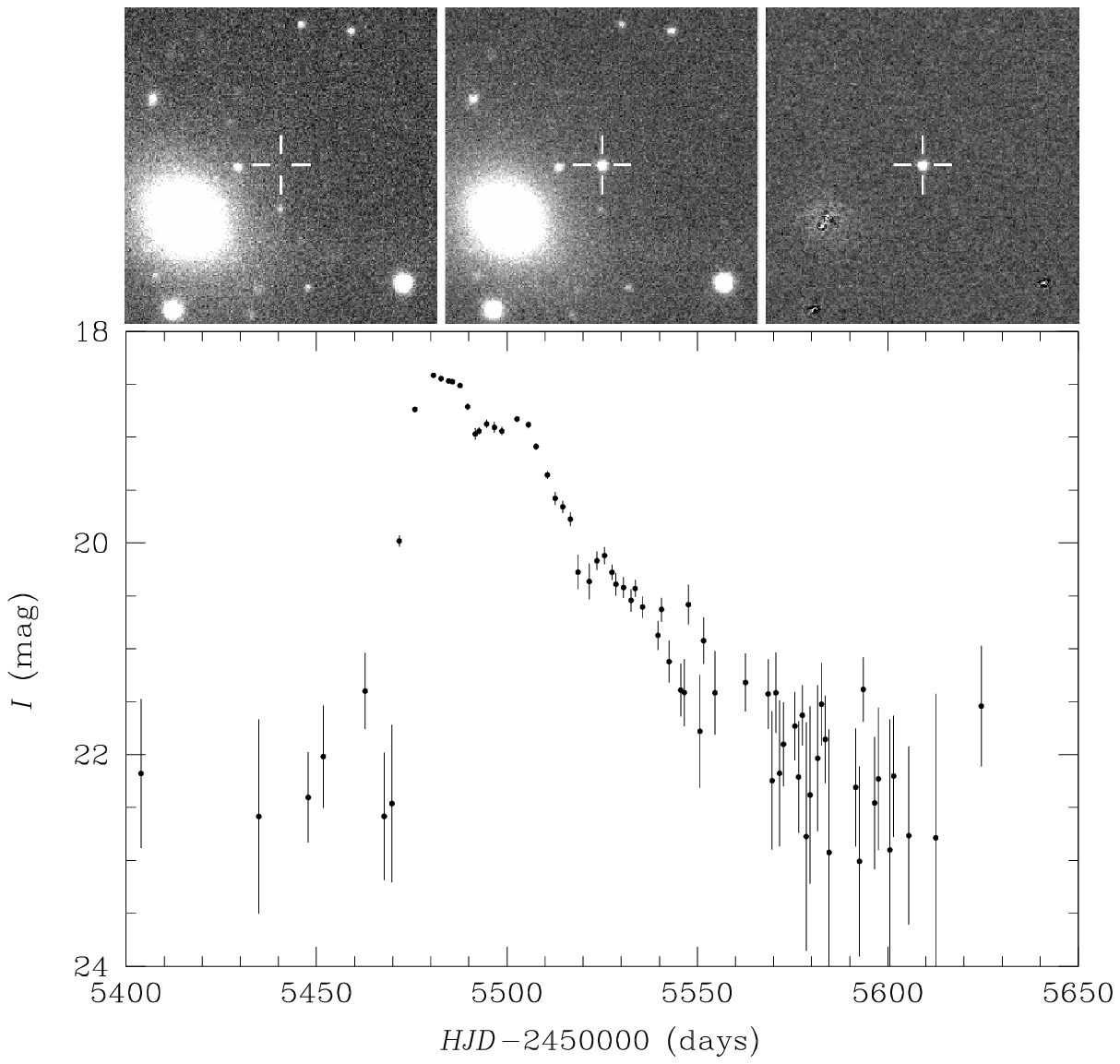}}
\vspace*{3mm}
\FigCap{{\it Top:} Finding chart for supernova OGLE-2010-SN-037 
(MBR108.29.613). Each image covers $60\arcs\times60\arcs$. The {\it left
image} shows the galaxy before the SN explosion, the {\it middle image}
shows the SN at its peak (marked with cross hair), and the {\it right
image} shows the difference image. The SN is located 21\zdot\arcs5 away
from the elliptical galaxy 2MASX~J02064540--7246329 at $z=0.057$. {\it
Bottom:} OGLE-IV light curve for OGLE-2010-SN-037. It peaked at
$I=18.41$~mag on 2010, October~11. The estimated absolute magnitude peak
was $M_I\approx-18.7$~mag. Both the shape of the light curve and the peak
absolute magnitude point to the SN Type Ia.}
\end{figure}
To search for SNe in the standard database, we adopted several parts
of the method used in Wyrzykowski \etal (2009). For each epoch with
magnitude $I_i\pm \Delta I_i$, we compute the significance of its
variability by
$$\sigma_i=\frac{I_{\rm med, B}-I_i}{\sqrt{\Delta I_i^2+\sigma_{\rm B}^2}}\eqno(1)$$
with respect to the variability in an outer window B, spanning half of
the data, and window A centered on the analyzed epoch. Here $I_{\rm
med, B}$ and $\sigma_{\rm B}$ are the median and dispersion calculated
in window B. We then searched for ``bumps'' defined as four (three)
consecutive data points with significance higher than 1.5 (1.8). The
63\,201 light curves showing such bumps were then inspected visually.
In the database of new objects, there were 8677 objects with at least
three detections. We produced full light curves at the locations of
these objects and then inspected them visually. We also carefully
checked the template, original and subtracted images around peak and
at the baseline.

\begin{figure}[htb]
\centerline{\includegraphics[width=11cm]{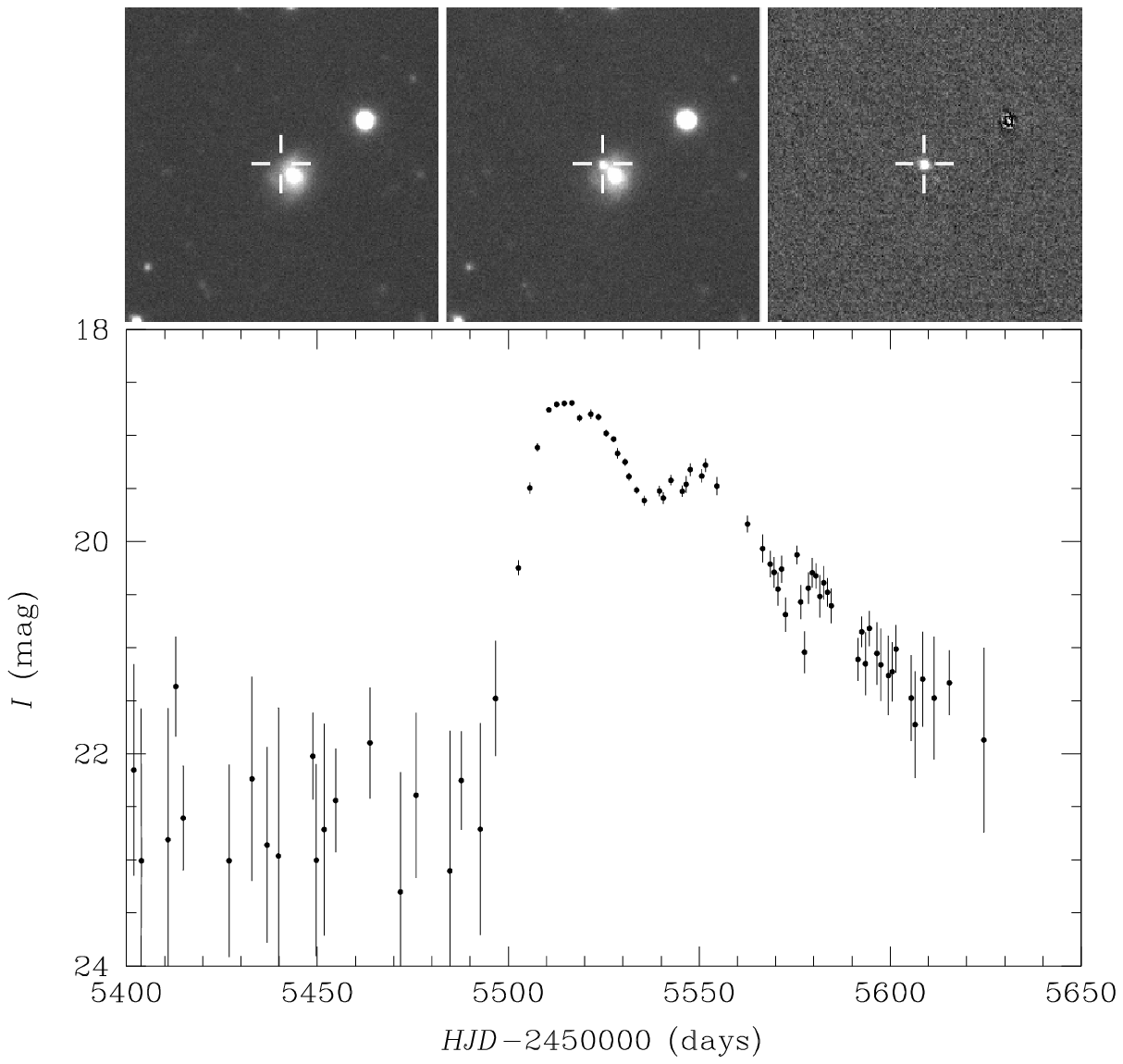}}
\vspace*{3mm}
\FigCap{{\it Top:} Finding chart for supernova OGLE-2010-SN-060
(MBR104.04.221). The {\it left image} shows the galaxy before the SN
explosion, the {\it middle image} shows the SN at its peak (marked with
cross hairs), and the {\it right image} shows the difference image. The SN
is located 3\zdot\arcs0 away from a spiral galaxy. The image covers
$60\arcs\times60\arcs$. {\it Bottom:} OGLE-IV light curve for
OGLE-2010-SN-060. It peaked at $I=18.69$~mag on 2010, November~14. The
shape of the light curve resembles that of SN Type Ia.}
\end{figure}
In this study we were interested in finding ``all plausible'' SNe and
transients that would allow us to calibrate our future automatic or
semi-automatic pipelines. We found 116 transients (three in overlapping
fields) in the standard database and 20 in the new database, of which three
were present in the standard database at slightly shifted positions.  Often
an extended galaxy is split into many ``point'' sources and the new
detection is too far ($>0\zdot\arcs5$) from any of the ``point'' sources to
be recognized as an object from the standard database. Such an exploding SN
then affects several nearby sources from the standard database leading to
``ghost'' variables adjacent to the real variable source. There were 249
such ghost variables. The real variables were recognized by their location
on the difference images. After inspecting the light curves we were left
with 128 plausible SNe and two foreground dwarf novae (Table~1) based on
the light curve shapes. Two spectacular SNe are presented in Figs.~3 and 4,
while light curves for basic SN types and two dwarf novae are shown in
Fig.~5. Out of 128 galaxies hosting plausible SNe, four have mid-infrared
colors consistent with an AGN (see Section~4 for details). These four
transients are faint, but two light curves seem to be consistent with SNe
and have flat light curves in the present 2012/2013 OGLE-IV season, so we
left them on our list of SNe. We removed from that list the two objects
that showed further variability in 2012. The remaining light curves were
mainly artifacts created by bright stars (moving/rotating spikes), other
artifacts associated with variable stars or were caused by photometric
problems.
\renewcommand{\arraystretch}{0.9}
\MakeTableSepp{c@{\hspace{7pt}}l@{\hspace{5pt}}c@{\hspace{5pt}}c@{\hspace{5pt}}c@{\hspace{5pt}}c@{\hspace{1pt}}r}{12.5cm}{OGLE-IV Transients from the OGLE-IV Magellanic Bridge Data}
{\hline
\noalign{\vskip3pt}
ID &\multicolumn{1}{c}{OGLE-IV} &    RA    &   DEC    & $T_{\rm  max}$ & $I_{\rm max}^*$  & Rem.\\
   &\multicolumn{1}{c}{Object No} & J2000.0 & J2000.0 & HJD [days] & [mag]         & \\
\noalign{\vskip3pt}
\hline
\noalign{\vskip3pt}
OGLE-2010-SN-002 & MBR102.19.2254   & 01\uph54\upm28\zdot\ups68 & $-72\arcd17\arcm09\zdot\arcs5$ & 245 5378 & 19.77 &  G  \\ 
OGLE-2010-SN-003 & MBR110.22.1291   & 02\uph04\upm01\zdot\ups45 & $-75\arcd21\arcm08\zdot\arcs2$ & 245 5380 & 20.12 &  G  \\ 
OGLE-2010-SN-004 & MBR112.03.727    & 02\uph24\upm29\zdot\ups34 & $-72\arcd57\arcm51\zdot\arcs3$ & 245 5387 & 20.38 &  --  \\ 
OGLE-2010-SN-005 & MBR128.28.1045   & 03\uph11\upm25\zdot\ups82 & $-72\arcd32\arcm31\zdot\arcs8$ & 245 5391 & 18.68 &  G  \\ 
OGLE-2010-SN-006 & MBR111.14.313    & 02\uph16\upm14\zdot\ups06 & $-71\arcd28\arcm23\zdot\arcs7$ & 245 5404 & 20.34 &  A  \\ 
OGLE-2010-SN-007 & MBR125.01.1359   & 03\uph10\upm39\zdot\ups36 & $-76\arcd38\arcm50\zdot\arcs9$ & 245 5404 & 20.37 &  --  \\ 
OGLE-2010-SN-008 & MBR144.22.109    & 03\uph58\upm37\zdot\ups67 & $-74\arcd59\arcm20\zdot\arcs6$ & 245 5404 & 17.84 &  G  \\ 
OGLE-2010-SN-009 & MBR118.29.422    & 02\uph38\upm26\zdot\ups47 & $-72\arcd41\arcm02\zdot\arcs4$ & 245 5405 & 19.92 &  G  \\ 
OGLE-2010-SN-010 & MBR122.13.1895   & 02\uph50\upm52\zdot\ups18 & $-72\arcd35\arcm12\zdot\arcs5$ & 245 5410 & 20.54 &  --  \\ 
OGLE-2010-SN-011 & MBR112.20.302    & 02\uph23\upm08\zdot\ups85 & $-72\arcd27\arcm20\zdot\arcs8$ & 245 5413 & 19.71 &  G  \\ 
OGLE-2010-SN-012 & MBR106.04.582    & 02\uph06\upm12\zdot\ups26 & $-71\arcd09\arcm50\zdot\arcs1$ & 245 5415 & 19.66 &  G  \\ 
OGLE-2010-SN-013 & MBR115.29.673    & 02\uph25\upm25\zdot\ups19 & $-75\arcd49\arcm02\zdot\arcs4$ & 245 5415 & 20.52 &  --  \\ 
OGLE-2010-SN-014 & MBR109.31.2418   & 02\uph00\upm52\zdot\ups95 & $-73\arcd48\arcm13\zdot\arcs1$ & 245 5427 & 19.60 &  G  \\ 
OGLE-2010-SN-015 & MBR142.18.476    & 03\uph57\upm18\zdot\ups30 & $-72\arcd29\arcm51\zdot\arcs9$ & 245 5431 & 20.53 &  --  \\ 
OGLE-2010-SN-016 & MBR125.26.1840   & 03\uph10\upm47\zdot\ups45 & $-75\arcd43\arcm38\zdot\arcs6$ & 245 5435 & 21.11 &  --  \\ 
OGLE-2010-SN-017 & MBR110.08.1348   & 02\uph16\upm55\zdot\ups45 & $-75\arcd37\arcm04\zdot\arcs9$ & 245 5436 & 21.14 &  G  \\ 
OGLE-2010-SN-018 & MBR135.03.572    & 03\uph39\upm40\zdot\ups72 & $-76\arcd43\arcm06\zdot\arcs0$ & 245 5436 & 20.25 &  --  \\ 
OGLE-2010-SN-019 & MBR113.05.899    & 02\uph21\upm36\zdot\ups62 & $-74\arcd11\arcm30\zdot\arcs2$ & 245 5438 & 20.23 &  G  \\ 
OGLE-2010-SN-020 & MBR131.08.1684   & 03\uph27\upm34\zdot\ups72 & $-71\arcd18\arcm44\zdot\arcs2$ & 245 5444 & 20.11 &  G  \\ 
OGLE-2010-SN-021 & MBR102.07.858    & 01\uph44\upm43\zdot\ups87 & $-73\arcd03\arcm36\zdot\arcs5$ & 245 5445 & 19.52 &  G  \\ 
OGLE-2010-SN-022 & MBR108.15.690    & 01\uph59\upm57\zdot\ups91 & $-73\arcd23\arcm34\zdot\arcs0$ & 245 5445 & 21.11 &  --  \\ 
OGLE-2010-SN-023 & MBR115.04.680    & 02\uph24\upm59\zdot\ups23 & $-76\arcd47\arcm15\zdot\arcs9$ & 245 5446 & 20.29 &  G  \\ 
OGLE-2010-SN-024 & MBR121.21.889    & 02\uph50\upm29\zdot\ups28 & $-71\arcd01\arcm31\zdot\arcs0$ & 245 5446 & 20.00 &  G  \\ 
OGLE-2010-SN-025 & MBR142.03.946    & 03\uph53\upm50\zdot\ups18 & $-72\arcd57\arcm50\zdot\arcs4$ & 245 5449 & 18.36 &  --  \\ 
OGLE-2010-SN-026 & MBR113.02.1083   & 02\uph28\upm08\zdot\ups57 & $-74\arcd05\arcm12\zdot\arcs1$ & 245 5450 & 19.24 &  G  \\ 
OGLE-2010-SN-027 & MBR121.32.1621   & 02\uph44\upm38\zdot\ups05 & $-70\arcd42\arcm39\zdot\arcs7$ & 245 5450 & 20.23 &  --  \\ 
OGLE-2010-SN-028 & MBR125.30.1229   & 03\uph00\upm36\zdot\ups10 & $-75\arcd37\arcm36\zdot\arcs5$ & 245 5450 & 20.23 &  G  \\ 
OGLE-2010-SN-029 & MBR103.06.2532   & 01\uph43\upm44\zdot\ups64 & $-74\arcd04\arcm44\zdot\arcs7$ & 245 5452 & 19.38 &  G  \\ 
OGLE-2010-SN-030 & MBR117.30.14N    & 02\uph35\upm04\zdot\ups98 & $-71\arcd28\arcm45\zdot\arcs8$ & 245 5462 & 19.36 &  G  \\ 
OGLE-2010-SN-031 & MBR136.27.269    & 03\uph39\upm20\zdot\ups40 & $-70\arcd15\arcm14\zdot\arcs2$ & 245 5465 & 19.95 &  G  \\ 
OGLE-2010-SN-032 & MBR133.26.541    & 03\uph31\upm45\zdot\ups13 & $-73\arcd19\arcm29\zdot\arcs0$ & 245 5468 & 20.62 &  --  \\ 
OGLE-2010-SN-033 & MBR118.27.32N    & 02\uph42\upm30\zdot\ups47 & $-72\arcd31\arcm23\zdot\arcs4$ & 245 5469 & 19.17 &  G  \\ 
OGLE-2010-SN-034 & MBR109.25.2848   & 01\uph56\upm44\zdot\ups15 & $-74\arcd07\arcm16\zdot\arcs4$ & 245 5470 & 20.10 &  G  \\ 
OGLE-2010-SN-035 & MBR111.07.1246   & 02\uph14\upm38\zdot\ups79 & $-71\arcd41\arcm11\zdot\arcs8$ & 245 5471 & 20.57 &  --  \\ 
OGLE-2010-SN-036 & MBR117.21.816    & 02\uph35\upm27\zdot\ups64 & $-71\arcd43\arcm29\zdot\arcs1$ & 245 5476 & 20.36 &  G  \\ 
OGLE-2010-SN-037 & MBR108.29.613    & 02\uph06\upm41\zdot\ups18 & $-72\arcd46\arcm22\zdot\arcs4$ & 245 5481 & 18.41 &  G  \\ 
OGLE-2010-SN-038 & MBR115.06.597    & 02\uph19\upm15\zdot\ups00 & $-76\arcd46\arcm56\zdot\arcs5$ & 245 5481 & 19.60 &  G  \\ 
OGLE-2010-SN-039 & MBR122.08.1171   & 03\uph00\upm49\zdot\ups14 & $-72\arcd34\arcm35\zdot\arcs9$ & 245 5481 & 19.39 &  --  \\ 
OGLE-2010-SN-040 & MBR145.22.290    & 04\uph03\upm44\zdot\ups04 & $-71\arcd54\arcm42\zdot\arcs7$ & 245 5484 & 19.62 &  G  \\ 
OGLE-2010-SN-041 & MBR138.26.198    & 03\uph44\upm25\zdot\ups87 & $-72\arcd40\arcm30\zdot\arcs3$ & 245 5485 & 20.23 &  G  \\ 
OGLE-2010-SN-042 & MBR142.15.1583   & 03\uph45\upm16\zdot\ups38 & $-72\arcd38\arcm50\zdot\arcs0$ & 245 5485 & 19.35 &  G  \\ 
OGLE-2010-SN-043 & MBR144.27.1885   & 04\uph06\upm36\zdot\ups14 & $-74\arcd25\arcm20\zdot\arcs6$ & 245 5485 & 20.43 &  --  \\ 
OGLE-2010-SN-044 & MBR123.27.710    & 02\uph59\upm44\zdot\ups89 & $-73\arcd14\arcm50\zdot\arcs2$ & 245 5491 & 17.29 &  G  \\ 
OGLE-2010-SN-045 & MBR130.18.1249   & 03\uph25\upm41\zdot\ups43 & $-75\arcd21\arcm51\zdot\arcs1$ & 245 5492 & 20.63 &  G  \\ 
OGLE-2010-SN-046 & MBR102.21.2068   & 01\uph49\upm26\zdot\ups92 & $-72\arcd17\arcm38\zdot\arcs9$ & 245 5493 & 19.97 &  G  \\ 
OGLE-2010-SN-047 & MBR136.26.177    & 03\uph40\upm55\zdot\ups79 & $-70\arcd18\arcm18\zdot\arcs1$ & 245 5494 & 19.76 &  G  \\ 
\hline
}
\setcounter{table}{0}
\MakeTableSepp{c@{\hspace{7pt}}l@{\hspace{5pt}}c@{\hspace{5pt}}c@{\hspace{5pt}}c@{\hspace{5pt}}c@{\hspace{1pt}}r}{12.5cm}{Continued}
{\hline
\noalign{\vskip3pt}
ID & \multicolumn{1}{c}{OGLE-IV} &    RA    &   DEC    & $T_{\rm  max}$ & $I_{\rm max}^*$                & Rem.\\
   &\multicolumn{1}{c}{Object No} & J2000.0 & J2000.0 & HJD [days] & [mag]         & \\
\noalign{\vskip3pt}
\hline
\noalign{\vskip3pt}
OGLE-2010-SN-048 & MBR100.04.859    & 01\uph52\upm17\zdot\ups85 & $-70\arcd36\arcm57\zdot\arcs1$ & 245 5497 & 20.18 &  G  \\ 
OGLE-2010-SN-049 & MBR142.28.339    & 03\uph53\upm13\zdot\ups28 & $-72\arcd11\arcm36\zdot\arcs3$ & 245 5498 & 19.95 &  G  \\ 
OGLE-2010-SN-050 & MBR141.25.790    & 03\uph43\upm24\zdot\ups65 & $-71\arcd05\arcm11\zdot\arcs9$ & 245 5500 & 20.51 &  G  \\ 
OGLE-2010-SN-051 & MBR135.30.1493   & 03\uph36\upm33\zdot\ups32 & $-75\arcd39\arcm37\zdot\arcs0$ & 245 5503 & 20.54 &  --  \\ 
OGLE-2010-SN-052 & MBR143.14.123    & 03\uph50\upm45\zdot\ups59 & $-74\arcd02\arcm14\zdot\arcs2$ & 245 5504 & 19.61 &  G  \\ 
OGLE-2010-SN-053 & MBR123.11.143    & 02\uph58\upm00\zdot\ups77 & $-74\arcd01\arcm59\zdot\arcs6$ & 245 5507 & 18.73 &  --  \\ 
OGLE-2010-SN-054 & MBR117.28.373    & 02\uph38\upm28\zdot\ups17 & $-71\arcd29\arcm21\zdot\arcs6$ & 245 5508 & 21.32 &  G  \\ 
OGLE-2010-SN-055 & MBR136.03.1063   & 03\uph37\upm55\zdot\ups82 & $-71\arcd03\arcm29\zdot\arcs3$ & 245 5508 & 19.61 &  G  \\ 
OGLE-2010-SN-056 & MBR119.27.1327   & 02\uph44\upm27\zdot\ups13 & $-73\arcd46\arcm13\zdot\arcs8$ & 245 5509 & 19.81 &  G  \\ 
OGLE-2010-SN-057 & MBR104.30.352    & 01\uph46\upm19\zdot\ups21 & $-74\arcd33\arcm21\zdot\arcs5$ & 245 5510 & 19.57 &  --  \\ 
OGLE-2010-SN-058 & MBR131.02.538    & 03\uph25\upm13\zdot\ups62 & $-71\arcd50\arcm33\zdot\arcs1$ & 245 5515 & 20.59 &  --  \\ 
OGLE-2010-SN-059 & MBR127.11.262    & 03\uph07\upm59\zdot\ups86 & $-72\arcd05\arcm31\zdot\arcs2$ & 245 5516 & 18.44 &  G  \\ 
OGLE-2010-SN-060 & MBR104.04.221    & 01\uph48\upm24\zdot\ups30 & $-75\arcd32\arcm47\zdot\arcs3$ & 245 5517 & 18.69 &  G  \\ 
OGLE-2010-SN-061 & MBR106.31.909    & 02\uph01\upm52\zdot\ups33 & $-70\arcd04\arcm44\zdot\arcs3$ & 245 5526 & 19.71 &  G  \\ 
OGLE-2010-SN-062 & MBR108.27.33N    & 02\uph10\upm18\zdot\ups20 & $-72\arcd43\arcm26\zdot\arcs9$ & 245 5531 & 17.61 &  --  \\ 
OGLE-2010-SN-063 & MBR119.02.638    & 02\uph44\upm52\zdot\ups44 & $-74\arcd51\arcm32\zdot\arcs9$ & 245 5533 & 19.98 &  G  \\ 
OGLE-2010-SN-064 & MBR106.06.328    & 02\uph01\upm17\zdot\ups18 & $-71\arcd16\arcm44\zdot\arcs6$ & 245 5534 & 20.42 &  G  \\ 
OGLE-2010-SN-065 & MBR144.26.1005   & 04\uph08\upm12\zdot\ups30 & $-74\arcd28\arcm12\zdot\arcs0$ & 245 5545 & 19.42 &  G  \\ 
OGLE-2010-SN-066 & MBR131.05.1464   & 03\uph18\upm01\zdot\ups95 & $-71\arcd43\arcm20\zdot\arcs1$ & 245 5547 & 19.78 &  --  \\ 
OGLE-2010-SN-067 & MBR100.03.953    & 01\uph52\upm47\zdot\ups44 & $-70\arcd33\arcm34\zdot\arcs4$ & 245 5548 & 20.49 &  G  \\ 
OGLE-2010-SN-068 & MBR116.13.532    & 02\uph34\upm10\zdot\ups79 & $-70\arcd52\arcm16\zdot\arcs2$ & 245 5550 & 18.54 &  G  \\ 
OGLE-2010-SN-069 & MBR119.30.63N    & 02\uph39\upm27\zdot\ups11 & $-73\arcd49\arcm23\zdot\arcs0$ & 245 5552 & 19.57 &  G  \\ 
OGLE-2010-SN-070 & MBR137.27.698    & 03\uph40\upm05\zdot\ups51 & $-71\arcd26\arcm13\zdot\arcs6$ & 245 5554 & 19.61 &  G  \\ 
OGLE-2010-SN-071 & MBR127.23.18N    & 03\uph02\upm23\zdot\ups69 & $-71\arcd44\arcm14\zdot\arcs4$ & 245 5557 & 18.80 &  G  \\ 
OGLE-2010-SN-072 & MBR131.28.564    & 03\uph22\upm49\zdot\ups56 & $-70\arcd50\arcm29\zdot\arcs5$ & 245 5557 & 20.07 &  --  \\ 
\hline
OGLE-2011-SN-007 & MBR101.23.216    & 01\uph47\upm38\zdot\ups12 & $-71\arcd14\arcm57\zdot\arcs1$ & 245 5563 & 19.42 &  G  \\ 
OGLE-2011-SN-008 & MBR135.32.2073N   & 03\uph31\upm31\zdot\ups90 & $-75\arcd47\arcm42\zdot\arcs9$ & 245 5579 & 19.13 &  G  \\ 
OGLE-2011-SN-009 & MBR123.32.2468N   & 02\uph49\upm26\zdot\ups55 & $-73\arcd22\arcm07\zdot\arcs6$ & 245 5582 & 17.80 &  G  \\ 
OGLE-2011-SN-010 & MBR129.25.1197N   & 03\uph02\upm59\zdot\ups96 & $-74\arcd20\arcm31\zdot\arcs9$ & 245 5592 & 18.49 &  G  \\ 
OGLE-2011-SN-011 & MBR109.02.53N    & 02\uph11\upm44\zdot\ups30 & $-74\arcd47\arcm20\zdot\arcs3$ & 245 5593 & 17.67 &  G  \\ 
OGLE-2011-SN-012 & MBR138.04.49N    & 03\uph38\upm24\zdot\ups06 & $-73\arcd35\arcm19\zdot\arcs1$ & 245 5596 & 18.77 &  G  \\ 
OGLE-2011-SN-013 & MBR115.24.75N    & 02\uph18\upm53\zdot\ups73 & $-76\arcd03\arcm06\zdot\arcs0$ & 245 5608 & 18.39 &  G  \\ 
OGLE-2011-SN-014 & MBR100.30.381    & 01\uph49\upm58\zdot\ups97 & $-69\arcd40\arcm13\zdot\arcs0$ & 245 5760 & 20.08 &  G  \\ 
OGLE-2011-SN-015 & MBR112.29.656    & 02\uph22\upm46\zdot\ups39 & $-71\arcd59\arcm15\zdot\arcs4$ & 245 5760 & 19.88 &  G  \\ 
OGLE-2011-SN-016 & MBR121.06.52N    & 02\uph47\upm17\zdot\ups33 & $-71\arcd49\arcm09\zdot\arcs8$ & 245 5761 & 18.58 &  --  \\ 
OGLE-2011-SN-017 & MBR127.17.641    & 03\uph13\upm27\zdot\ups68 & $-71\arcd48\arcm48\zdot\arcs5$ & 245 5785 & 19.13 &  G  \\ 
OGLE-2011-SN-018 & MBR123.22.567N   & 02\uph53\upm17\zdot\ups38 & $-73\arcd42\arcm54\zdot\arcs5$ & 245 5788 & 18.50 &  G  \\ 
OGLE-2011-SN-019 & MBR121.07.686    & 02\uph45\upm34\zdot\ups06 & $-71\arcd48\arcm25\zdot\arcs7$ & 245 5799 & 19.97 &  G  \\ 
OGLE-2011-SN-020 & MBR133.24.403    & 03\uph19\upm09\zdot\ups15 & $-73\arcd43\arcm36\zdot\arcs8$ & 245 5804 & 21.28 &  G  \\ 
OGLE-2011-SN-021 & MBR141.24.965    & 03\uph44\upm06\zdot\ups00 & $-71\arcd07\arcm56\zdot\arcs4$ & 245 5806 & 19.88 &  G  \\ 
OGLE-2011-SN-022 & MBR114.20.214    & 02\uph27\upm38\zdot\ups11 & $-74\arcd56\arcm55\zdot\arcs0$ & 245 5808 & 20.30 &  G A \\ 
OGLE-2011-SN-023 & MBR116.17.1214   & 02\uph42\upm59\zdot\ups19 & $-70\arcd27\arcm39\zdot\arcs0$ & 245 5811 & 20.30 &  --  \\ 
OGLE-2011-SN-024 & MBR110.05.116    & 02\uph04\upm10\zdot\ups92 & $-76\arcd10\arcm22\zdot\arcs3$ & 245 5812 & 20.37 &  G  \\ 
OGLE-2011-SN-025 & MBR117.22.662    & 02\uph33\upm32\zdot\ups70 & $-71\arcd46\arcm13\zdot\arcs1$ & 245 5812 & 20.36 &  G  \\ 
OGLE-2011-SN-026 & MBR126.30.875    & 03\uph03\upm25\zdot\ups99 & $-70\arcd05\arcm25\zdot\arcs0$ & 245 5812 & 19.81 &  G  \\ 
OGLE-2011-SN-027 & MBR130.05.1170   & 03\uph15\upm11\zdot\ups01 & $-76\arcd02\arcm45\zdot\arcs0$ & 245 5812 & 19.99 &  G  \\ 
\hline
}
\setcounter{table}{0}
\MakeTableSepp{c@{\hspace{7pt}}l@{\hspace{5pt}}c@{\hspace{5pt}}c@{\hspace{5pt}}c@{\hspace{5pt}}c@{\hspace{1pt}}r}{12.5cm}{Concluded}
{\hline
\noalign{\vskip3pt}
ID & \multicolumn{1}{c}{OGLE-IV} &    RA    &   DEC    & $T_{\rm  max}$ & $I_{\rm max}^*$                & Rem. \\
   &\multicolumn{1}{c}{Object No} & J2000.0 & J2000.0 & HJD [days] & [mag]         & \\
\noalign{\vskip3pt}
\hline
\noalign{\vskip3pt}
OGLE-2011-SN-028 & MBR130.01.1341   & 03\uph25\upm49\zdot\ups60 & $-76\arcd01\arcm41\zdot\arcs1$ & 245 5813 & 20.26 &  G  \\ 
OGLE-2011-SN-029 & MBR145.32.271    & 03\uph59\upm48\zdot\ups72 & $-71\arcd33\arcm27\zdot\arcs4$ & 245 5817 & 20.73 &  G  \\ 
OGLE-2011-SN-030 & MBR102.16.792    & 01\uph40\upm52\zdot\ups31 & $-72\arcd43\arcm18\zdot\arcs7$ & 245 5830 & 20.00 &  G  \\ 
OGLE-2011-SN-031 & MBR144.32.438    & 03\uph54\upm49\zdot\ups04 & $-74\arcd32\arcm55\zdot\arcs6$ & 245 5833 & 20.34 &  G  \\ 
OGLE-2011-SN-032 & MBR102.08.2592   & 01\uph58\upm11\zdot\ups17 & $-72\arcd32\arcm04\zdot\arcs0$ & 245 5839 & 20.41 &  G  \\ 
OGLE-2011-SN-033 & MBR131.30.1305   & 03\uph18\upm26\zdot\ups34 & $-70\arcd44\arcm07\zdot\arcs1$ & 245 5841 & 20.51 &  G  \\ 
OGLE-2011-SN-034 & MBR121.29.882    & 02\uph50\upm18\zdot\ups34 & $-70\arcd52\arcm25\zdot\arcs8$ & 245 5842 & 20.12 &  G  \\ 
OGLE-2011-SN-035 & MBR126.20.189    & 03\uph08\upm05\zdot\ups00 & $-70\arcd38\arcm26\zdot\arcs9$ & 245 5842 & 19.94 &  G  \\ 
OGLE-2011-SN-036 & MBR122.20.1109   & 02\uph53\upm23\zdot\ups21 & $-72\arcd19\arcm55\zdot\arcs7$ & 245 5845 & 19.64 &  G  \\ 
OGLE-2011-SN-037 & MBR123.02.1141   & 02\uph59\upm13\zdot\ups85 & $-74\arcd11\arcm35\zdot\arcs7$ & 245 5856 & 20.15 &  G  \\ 
OGLE-2011-SN-038 & MBR132.04.1169   & 03\uph21\upm24\zdot\ups93 & $-72\arcd56\arcm09\zdot\arcs5$ & 245 5863 & 20.30 &  --  \\ 
OGLE-2011-SN-039 & MBR137.13.246N   & 03\uph34\upm02\zdot\ups19 & $-72\arcd09\arcm25\zdot\arcs1$ & 245 5865 & 19.11 &  G  \\ 
OGLE-2011-SN-040 & MBR118.10.1378   & 02\uph41\upm49\zdot\ups51 & $-73\arcd09\arcm12\zdot\arcs2$ & 245 5866 & 20.31 &  G  \\ 
OGLE-2011-SN-041 & MBR107.04.1444   & 02\uph05\upm13\zdot\ups75 & $-72\arcd16\arcm28\zdot\arcs4$ & 245 5870 & 20.54 &  G  \\ 
OGLE-2011-SN-042 & MBR102.23.1484   & 01\uph45\upm17\zdot\ups90 & $-72\arcd15\arcm32\zdot\arcs6$ & 245 5871 & 20.09 &  G  \\ 
OGLE-2011-SN-043 & MBR119.03.957    & 02\uph43\upm47\zdot\ups68 & $-74\arcd43\arcm18\zdot\arcs6$ & 245 5871 & 18.03 &  G  \\ 
OGLE-2011-SN-044 & MBR106.29.211    & 02\uph06\upm31\zdot\ups94 & $-70\arcd17\arcm30\zdot\arcs8$ & 245 5874 & 20.12 &  G  \\ 
OGLE-2011-SN-045 & MBR126.30.805    & 03\uph03\upm06\zdot\ups86 & $-70\arcd10\arcm26\zdot\arcs3$ & 245 5874 & 19.63 &  G  \\ 
OGLE-2011-SN-046 & MBR138.23.558    & 03\uph33\upm19\zdot\ups54 & $-72\arcd55\arcm18\zdot\arcs8$ & 245 5874 & 19.83 &  G  \\ 
OGLE-2011-SN-047 & MBR102.02.911    & 01\uph53\upm29\zdot\ups05 & $-73\arcd06\arcm55\zdot\arcs1$ & 245 5879 & 19.87 &  G  \\ 
OGLE-2011-SN-048 & MBR128.30.1799   & 03\uph05\upm44\zdot\ups73 & $-72\arcd38\arcm51\zdot\arcs9$ & 245 5886 & 20.58 &  G  \\ 
OGLE-2011-SN-049 & MBR122.26.1156   & 02\uph59\upm04\zdot\ups44 & $-72\arcd00\arcm29\zdot\arcs0$ & 245 5890 & 19.09 &  G  \\ 
OGLE-2011-SN-050 & MBR134.04.906    & 03\uph31\upm27\zdot\ups83 & $-75\arcd24\arcm23\zdot\arcs1$ & 245 5906 & 18.80 &  G  \\ 
OGLE-2011-SN-051 & MBR115.13.276N   & 02\uph23\upm25\zdot\ups06 & $-76\arcd27\arcm53\zdot\arcs3$ & 245 5910 & 17.91 &  G  \\ 
OGLE-2011-SN-052 & MBR121.31.984    & 02\uph46\upm42\zdot\ups15 & $-70\arcd40\arcm18\zdot\arcs7$ & 245 5910 & 20.39 &  G  \\ 
OGLE-2011-SN-053 & MBR139.27.895    & 03\uph48\upm47\zdot\ups87 & $-73\arcd48\arcm08\zdot\arcs8$ & 245 5925 & 19.31 &  --  \\ 
\hline
OGLE-2012-SN-053 & MBR102.15.317    & 01\uph44\upm38\zdot\ups46 & $-72\arcd42\arcm46\zdot\arcs6$ & 245 5929 & 20.75 &  G  \\ 
OGLE-2012-SN-054 & MBR120.19.1297   & 02\uph49\upm20\zdot\ups06 & $-75\arcd24\arcm09\zdot\arcs3$ & 245 5938 & 19.94 &  G  \\ 
OGLE-2012-SN-055 & MBR127.02.1343   & 03\uph09\upm49\zdot\ups04 & $-72\arcd15\arcm18\zdot\arcs9$ & 245 5938 & 20.00 &  G  \\ 
OGLE-2012-SN-056 & MBR129.02.121    & 03\uph17\upm20\zdot\ups34 & $-74\arcd52\arcm10\zdot\arcs1$ & 245 5942 & 19.25 &  G  \\ 
OGLE-2012-SN-057 & MBR140.30.903    & 03\uph47\upm15\zdot\ups56 & $-75\arcd00\arcm05\zdot\arcs4$ & 245 5948 & 20.07 &  G  \\ 
OGLE-2012-SN-058 & MBR123.09.1095   & 03\uph03\upm01\zdot\ups93 & $-73\arcd51\arcm11\zdot\arcs4$ & 245 5964 & 19.87 &  --  \\ 
OGLE-2012-SN-059 & MBR102.06.2066   & 01\uph46\upm08\zdot\ups74 & $-72\arcd55\arcm43\zdot\arcs5$ & 245 5982 & 20.12 &  G  \\ 
OGLE-2012-SN-060 & MBR145.30.487    & 04\uph03\upm51\zdot\ups59 & $-71\arcd30\arcm42\zdot\arcs2$ & 245 5983 & 20.04 &  --  \\
\hline
Dwarf Nova & MBR124.21.11N          & 02\uph59\upm28\zdot\ups56 & $-74\arcd48\arcm25\zdot\arcs1$ & 245 5414 & 16.15   &  -- \\ 
Dwarf Nova & MBR120.21.138N         & 02\uph44\upm25\zdot\ups24 & $-75\arcd33\arcm29\zdot\arcs9$ & 245 5802 & 15.89   &  -- \\ 
AGN & MBR134.31.146                 & 03\uph26\upm47\zdot\ups14 & $-74\arcd37\arcm51\zdot\arcs8$ & 245 5459 & 19.08   &  G A \\
AGN & MBR140.05.1125                & 03\uph48\upm38\zdot\ups86 & $-76\arcd03\arcm36\zdot\arcs1$ & 245 5445 & 20.05   &  A \\
\hline
\noalign{\vskip9pt}
\multicolumn{7}{p{12.5cm}}{Notes. $^*$The peak SNe magnitudes were
corrected to show only SNe fluxes (the median baseline fluxes were
removed). In the first column (ID) the supposedly missing SN numbers at the
beginning of each year were already reported in other OGLE papers.  In the
second column (OGLE-IV Object No) objects ending with an ``N'' are
discovered in the database of new objects.  The remaining objects were
found in the standard database. In the last column (Remarks) ``G'' means
that the SN host galaxy is present on the template image ($I<21$~mag) and
``A'' means that the WISE $W1-W2$ color of the host galaxy is consistent
with an AGN (see Section~4).}}
\begin{figure}[htb]
\centerline{\includegraphics[width=11.9cm]{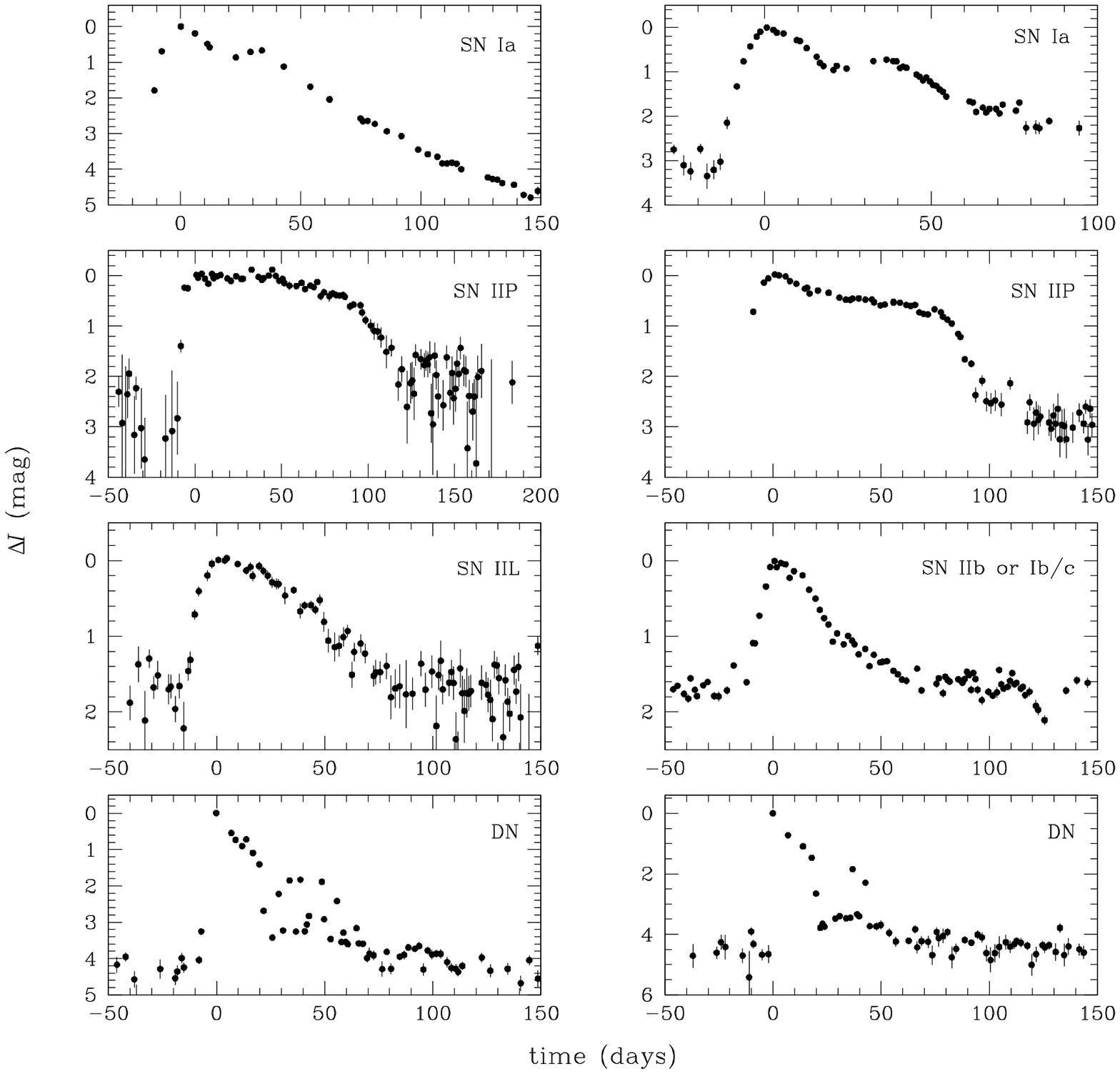}}
\vspace*{3mm}
\FigCap{OGLE-IV light curves for SNe ({\it three upper rows}) and dwarf
novae ({\it bottom row}). Plausible SN types are based on the light
curve shape only and are marked inside panels.}
\end{figure}

\Section{SN Numbers}
\vspace*{-3pt}
In this section, we estimate the number of expected SNe as a function
of magnitude (based on available SNe rates) and hence the OGLE-IV SN
detection efficiency. A source of absolute magnitude $M_I$ is observed
at magnitude
\vspace*{-3pt}
$$m_I=M_I+5\log{d(z,\Omega_X)}+25+A_{\rm Galaxy}+A_{\rm Host}(z)+K(z,~A_{\rm Host}(z))\eqno(2)$$ 
\vspace*{-3pt}
where $d(z, \Omega_X)$ is the luminosity distance in Mpc, a function
of redshift $z$ and cosmological model $\Omega_X$, $A_{\rm Galaxy}$
and $A_{\rm Host}(z)$ are the extinctions in our own Galaxy and in the
SN host galaxy, respectively, and $K(z,~A_{\rm Host}(z))$ is the
single filter {\it K}-correction (see Kim, Goobar and Perlmutter 1996
and Nugent, Kim and Perlmutter 2002). We used a standard $\Lambda$CDM
cosmological model with ($\Omega_\Lambda,\Omega_M,\Omega_k)=
(0.7,0.3,0.0)$ and $h=H_0/(100~{\rm km\,s^{-1}\,Mpc^{-1}})=0.73$ to calculate
distances. The Galactic extinction was set to $A_{\rm
Galaxy}=0.11$~mag (Schlegel, Finkbeiner and Davis 1998) and, for simplicity, the
median SNe Type Ia and ccSNe host extinctions were assumed to be
$A_{\rm Host~SNIa}=0.08$~mag and $A_{\rm Host~ccSNe}=0.22$~mag at
$z=0$ (\eg Holwerda 2008 and Schmidt \etal 1994, respectively). The
host extinctions at different redshifts in filter {\it I} were
calculated using a standard extinction law with $R_V=3.1$. To
calculate {\it K}-corrections (Fig.~6), we used the SN spectral
templates for the peak brightness from Peter Nugent's spectral
templates website\footnote{\it
http://supernova.lbl.gov/$\sim$nugent/nugent\_templates.html} and the
formula for the single filter {\it K}-correction from Kim, Goobar and
Perlmutter (1996; their Eq.~1). Because OGLE SNe light curves often
contain both flux from SNe and their host galaxies, the peak
magnitudes were corrected to show only SNe fluxes (median baseline
fluxes were removed).
\begin{figure}[htb]
\centerline{\includegraphics[width=11.8cm]{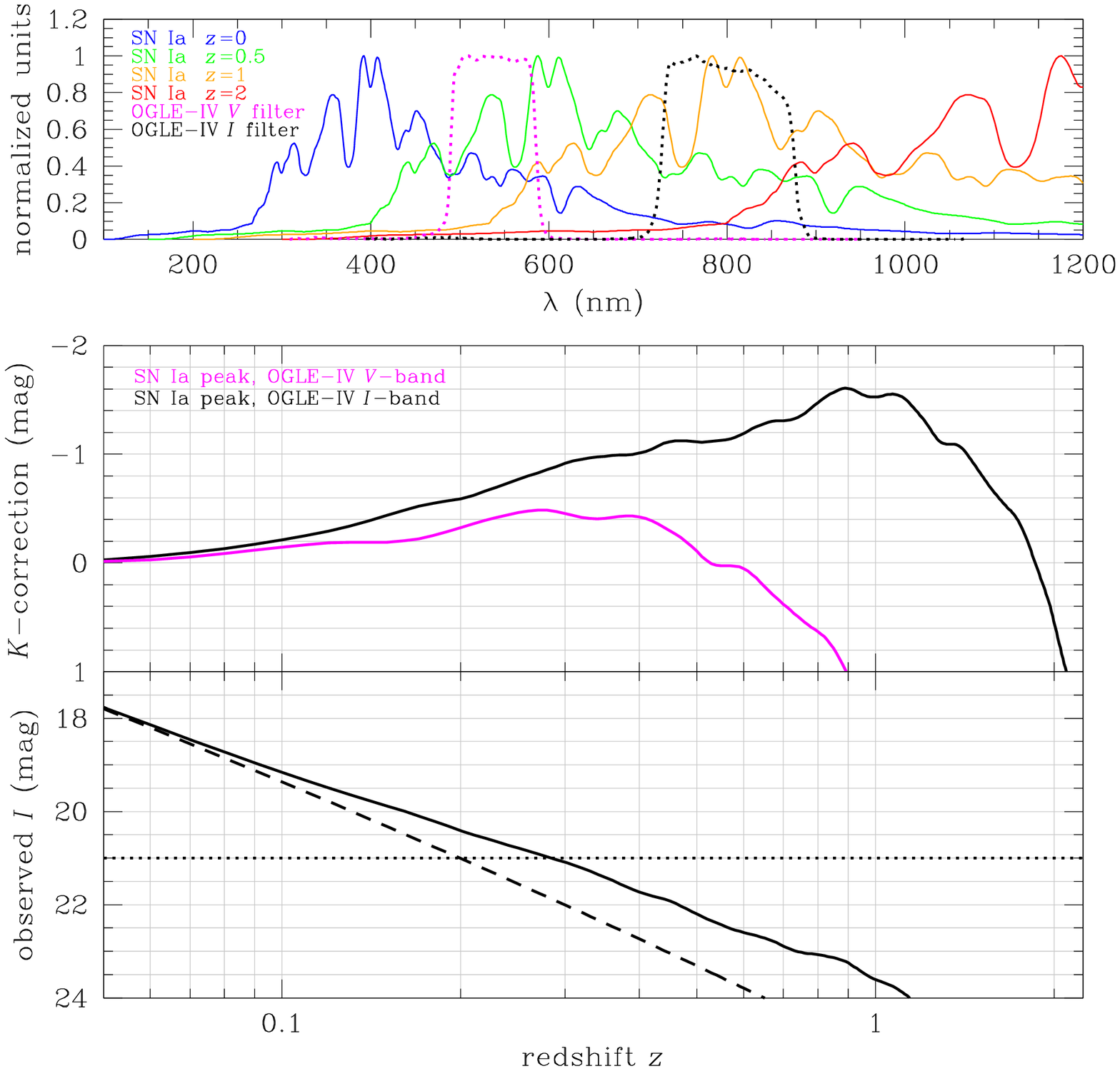}}
\vspace*{3mm}
\FigCap{Impact of redshift on the observed properties of SN Type~Ia. 
{\it Top:} SN Type~Ia peak spectrum ($t=0$ days) shown at four
redshifts $z=0$, 0.5, 1, and 2 (blue, green, yellow, and red lines,
respectively). We also show the OGLE-IV {\it V}- and {\it I}-band
filter transmission fractions convolved with the quantum efficiency of
the OGLE-IV CCDs (dotted magenta and black, respectively). Both the SN
spectra and filter transmissions are normalized to~1. {\it Middle:}
{\it K}-corrections for the SN Type Ia peak spectrum ($t=0$ days) as a
function of redshift for the {\it V}- and {\it I}-band filters (solid
magenta and black, respectively). The {\it I}-band absolute brightness
of SN Type Ia increases with increasing redshift, with {\it
K}-corrections peaking at $K\approx-1.6$~mag for $z\approx0.9$, and
then decreases when the spectrum peak leaves the {\it I}-band filter
($z>1$). {\it Bottom:} The observed {\it I}-band peak magnitudes as a
function of redshift (solid line). The dashed line is for the observed
{\it I}-band peak magnitudes with no {\it K}-corrections. The negative
{\it K}-correction increases the limiting redshift from 0.20 to 0.28
given the OGLE-IV magnitude limit of $I\approx21$~mag (horizontal
dotted line), tripling the effective survey volume.}
\end{figure}

\subsection{The Expected SN Numbers}
There are two ways to obtain the SNe detection efficiency for a survey.
The first one is to simulate SN light curves of various types with the
OGLE-IV cadence and photometric properties, and check if they pass our
``bump'' test (Section~2). Such a simulation is difficult to perform due to
many input unknowns and the interpretation may be ambiguous. While we are
interested in SNe light curves only, they often include flux coming from
host galaxies. It is unknown what fraction of light coming from a galaxy
should be added to the SN light, since some SNe explode in the outskirts of
galaxies some near centers, SNe span a wide range of brightness, and one
has to include extinctions. This would require simulating images of
galaxies with exploding SNe and then running through standard OGLE-IV
photometric pipelines. The second problem is that it is necessary to assume
some fractions of different types of SNe to a limiting magnitude (explained
in details in the second simulation) to calculate a proper SNe detection
efficiency of a survey. Therefore, we decided to perform a different
simulation, where we compare our number of SNe (their peak magnitudes) with
simulated number of SNe to a given limiting peak magnitude.

To calculate the expected number of SNe as a function of {\it I}-band
limiting peak magnitude (Fig.~7), we used the volumetric SN rates from
Li \etal (2011a).  For the SN Type Ia, Ibc, and II, we used SN rates
of
$0.301\times10^{-4}~{\rm SN\,yr^{-1}\,Mpc^{-3}}$,
$0.258\times10^{-4}~{\rm SN\,yr^{-1}\,Mpc^{-3}}$, and 
$0.447\times10^{-4}~{\rm SN\,yr^{-1}\,Mpc^{-3}}$, 
respectively, that evolve with redshift as $\propto(1+z)^{3.6}$. We
also used the fractions of various SNe within core-collapse group from
Li \etal (2011b), namely IIP (70\%), IIL (10\%), IIb (12\%), and IIn
(9\%).

The peak {\it I}-band magnitude for SN Type Ia was taken from Riess,
Press and Kirshner (1996) and was set to $M_I=-19.05$~mag. We
simplified our calculations by combining SNe Type Ib, Ic, and Ibc into
one group for which we set the peak magnitude to $M_I=-16.7$~mag,
converting $M_R$ from Li \etal (2011b). For SN Type II, we adopt
$M_I=-17.6$~mag, $-18.05$~mag, and $-17.4$~mag for SN Type IIP, IIL,
and IIb/IIn, respectively. The value for SN Type IIP is
consistent with Otsuka \etal (2012) and the value for SN Type IIb
consistent with Tsvetkov \etal (2012), SN Type IIn span a huge range
of absolute magnitudes ($-15$ to $-23$) and our choice differs from
the average of $M_I=-18.9$~mag adopted in Kiewe \etal (2012).
\begin{figure}[t]
\centerline{\includegraphics[width=11cm]{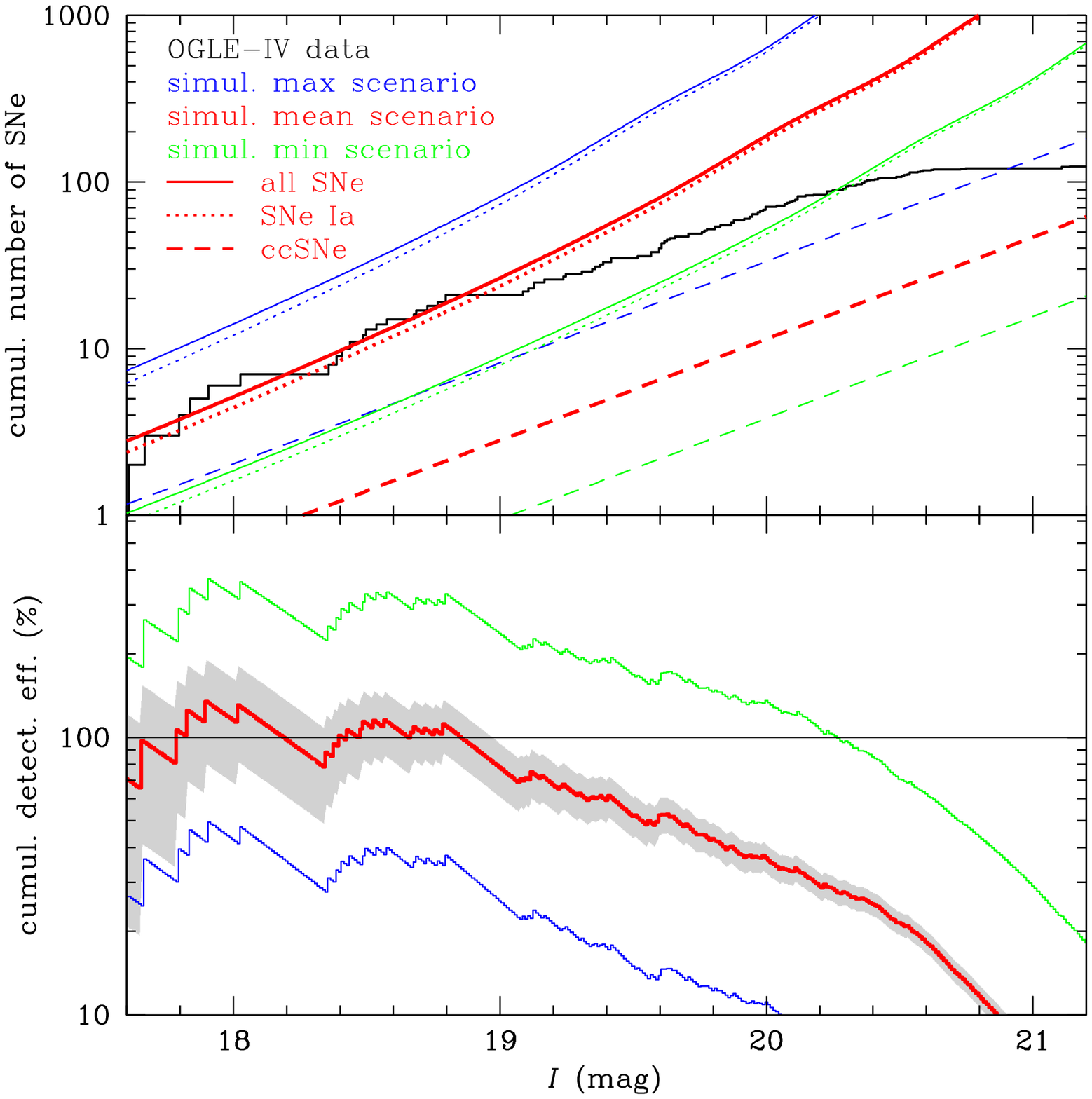}}
\vspace*{3mm}
\FigCap{{\it Top.} Cumulative number of SNe discovered by OGLE-IV 
as a function of the limiting peak {\it I}-band magnitude (black solid
line). Solid color lines show the total number of predicted SNe, while the
dotted line is for SNe Type Ia only and dashed line is for core-collapse
SNe. Red color marks the prediction for typical absolute peak magnitudes
and typical volumetric rates. Blue and green colors mark the optimistic
(absolute magnitudes brighter and volumetric SNe rates higher than in a
typical scenario) and pessimistic scenarios, respectively (see text for
details). {\it Bottom.} Cumulative detection efficiency of SNe as a
function of a limiting peak {\it I}-band magnitude. Line styles and color
coding is identical as in the {\it top panel}. The gray area reflects the
$1\sigma$ Poisson uncertainties for the typical scenario (red).}
\end{figure}

With the exception of SN Type Ia, the distribution of absolute
magnitudes for each SN type is not precisely known. There is no
evidence that they are Gaussian, so we estimated the maximum and
minimum number of expected SNe (hereafter referred to as the
optimistic and pessimistic scenarios), by taking into account the
uncertainties in rates and peak magnitudes. For SN Type Ia, we adopted
the range of {\it I}-band magnitudes of $\pm0.5$~mag, for Type Ibc
$\pm0.4$~mag, for Types IIP $\pm0.6$~mag, IIL $\pm0.3$~mag and IIb/n
$\pm0.8$~mag. For the optimistic (pessimistic) scenario, we simply
made all SNe brighter (fainter) by the corresponding magnitude shift
while simultaneously increased (decreased) the SN rates by their
reported uncertainties.

We analyzed two OGLE-IV MBR seasons. The seasonal gaps account for
35\% of time annually, so the ``OGLE-IV SNe MBR season length''
accounts for 65\% of the year. Our simulated SNe numbers are then
calculated for 1.3~year.

\renewcommand{\arraystretch}{1.1}
\MakeTableee{lccccccc}{12.5cm}
{Cumulative estimated number of SNe  per 100~deg$^2$ per year to the
limiting peak magnitude {\it I}}
{\hline
\noalign{\vskip3pt}
Peak & \multicolumn{3}{c}{Total SN Number} & Cumulative Detection & \multicolumn{3}{c}{OGLE-IV SN Number$^*$} \\
{\it I} [mag] & 
\multicolumn{1}{c}{All SNe} & 
\multicolumn{1}{c}{SNe Ia} & 
\multicolumn{1}{c}{ccSNe} & 
Efficiency (\%) &
\multicolumn{1}{c}{All SNe} & 
\multicolumn{1}{c}{SNe Ia} & 
\multicolumn{1}{c}{ccSNe} \\
\noalign{\vskip3pt}
\hline
\noalign{\vskip3pt}
17.0 &   1 &   1 &  0 & 100 &   1 &   1 &  0 \\
17.5 &   3 &   2 &  1 & 100 &   2 &   1 &  1 \\
18.0 &   6 &   5 &  1 & 100 &   4 &   3 &  1 \\
18.5 &  14 &  12 &  2 & 100 &   9 &   8 &  1 \\
19.0 &  31 &  28 &  3 &  83 &  17 &  15 &  2 \\
19.5 &  78 &  71 &  7 &  58 &  30 &  27 &  3 \\
20.0 & 224 & 211 & 13 &  38 &  56 &  53 &  3 \\
20.5 & 589 & 562 & 27 &  21 &  83 &  79 &  4 \\
\hline
\noalign{\vskip4pt}
\multicolumn{8}{p{11.5cm}}
{Notes. $^*$The reported SNe numbers are corrected for the OGLE-IV SNe
``season length'' ($\approx65\%$ of year).}}
We show the results of our simulations in Fig.~7 and Table~2. Our SNe
search is $\approx100\%$ complete for $I<18.8$~mag and drops to 50\% at
$I\approx19.7$~mag. In Table~2, we show the expected number of SNe per
100~deg$^2$ per year as a function of the limiting magnitude both for
the OGLE-IV survey setup and absolute numbers.

\Section{Discriminating SNe from AGNs}
AGNs are optically variable and increase our false-positive rate,
especially since we detect many ``transients'' near the centers of
galaxies. A key point is that bright and/or extended galaxies are
usually split into many ``point'' sources by our pipeline, so that
even for bright galaxies there may be no detection of an object at the
galaxy's center in the standard database and a transient can appear as
new object.

One of the key features of AGNs is their dust emission at mid-infrared
(mid-IR) wavelengths. The mid-IR colors of quasars are well established
from modern space missions (see Stern \etal 2005, Assef \etal 2010,
Koz³owski \etal 2010a, Stern \etal 2012) such as Spitzer Space
Telescope and Wide-field Infrared Survey Explorer (WISE, Wright \etal
2010). WISE observed the whole sky at 3.4~$\mu m$, 4.6~$\mu m$, 12~$\mu m$,
and 22~$\mu m$, hereafter called $W1$, $W2$, $W3$, and $W4$ bands,
respectively, and is an ideal source permitting removal of AGNs from our
sample. Stern \etal (2012) proposed a simple cut based on $W1$ and $W2$
bands, namely $W1-W2\geq0.8$~mag, that selects 78\% AGNs with 95\%
reliability for $W2\lesssim15.0$~mag. The remaining two bands are not
nearly as sensitive as the first two, and were not used.  Assef \etal
(2012), based on $\approx1000$~AGNs from the 9~deg$^2$ NOAO Bootes Field,
proposed another relation that goes 2~mag deeper than that of Stern \etal
(2012), $W1-W2>0.53\times\exp{(0.18\times(W2-13.76))}$~mag for
$W2<17.11$~mag that returns AGN with 75\% reliability. Their second and
more stringent relation (90\% reliability) is not used here, since we are
not interested in reliably identifying AGNs, but rather in discriminating
plausible AGNs.
\begin{figure}[htb]
\centerline{\includegraphics[width=12cm]{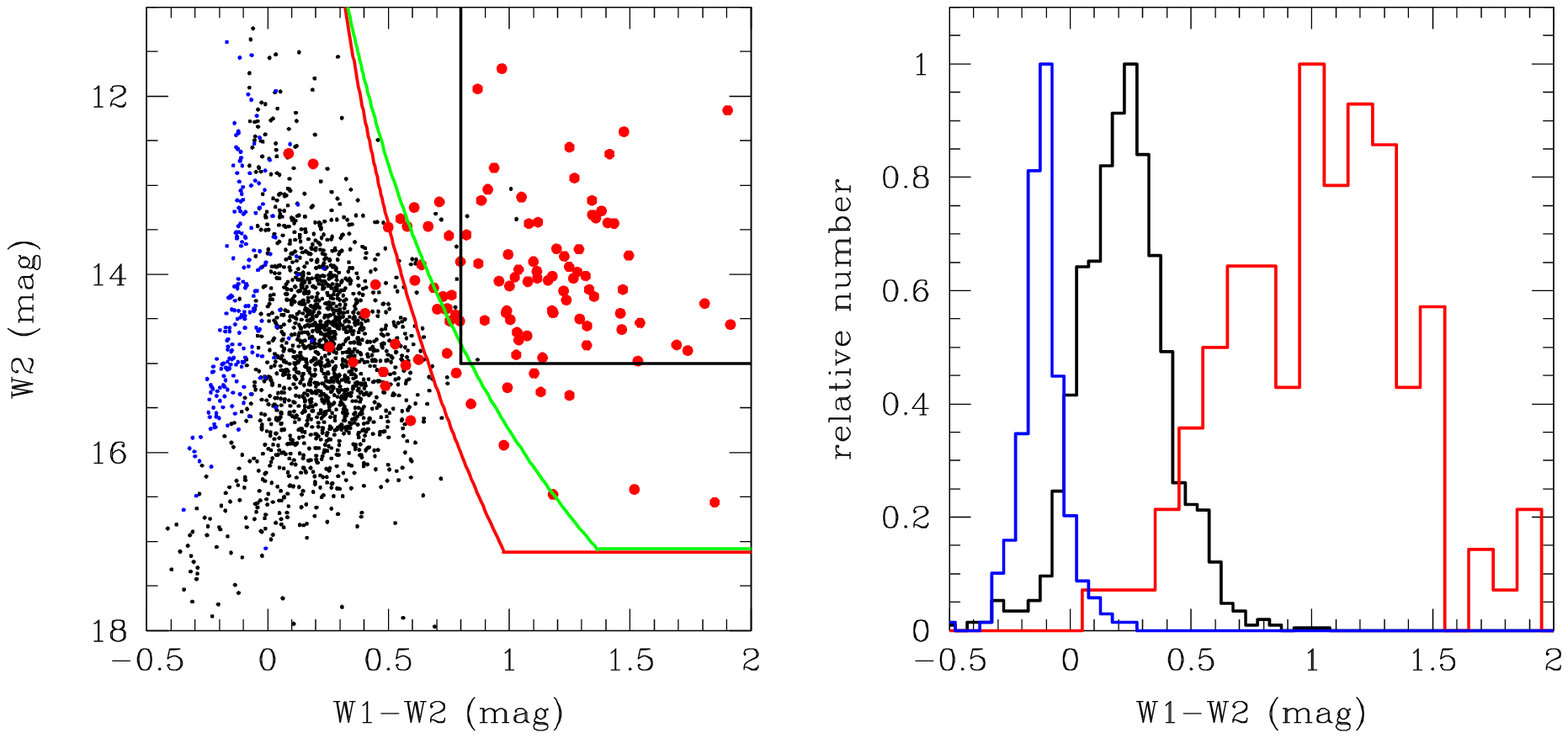}}
\vspace*{3mm}
\FigCap{{\it Left:} WISE color--magnitude diagram for a subset of the 
OGLE-IV GSEP stars (blue), galaxies (black), and OGLE-III LMC AGNs (red).
The black lines mark the Stern \etal (2012) AGN selection region, the
red (green) lines mark the Assef \etal (2012) AGN selection region for
75\% (90\%) reliability. Both selection methods identify the majority of
the known AGNs. {\it Right:} Normalized histograms for OGLE-IV GSEP
stars (blue), galaxies (black), and OGLE-III LMC AGNs (red). Median
WISE $W1-W2$ colors are $-0.12$~mag, 0.22~mag, and 1.07~mag,
respectively.}
\end{figure}

We matched the OGLE-III LMC confirmed quasars (Koz³owski \etal
2012), galaxies detected in the OGLE-IV GAIA South Ecliptic Pole
fields (GSEP, Soszyñ\-ski \etal 2012), and a subset of GSEP stars to
the WISE mid-IR database to verify these relations (Fig.~8).

Of the 169 LMC quasars, we matched 111 (66\%) to WISE (3\arcs matching
radius), where 98 meet the second criterion ($W2\lesssim15.0$~mag)
from Stern \etal (2012) and all of them are brighter than
$W2=17.11$~mag (second criterion from Assef \etal 2012). Both samples
span $0.09<W1-W2<1.92$~mag, with the medians $W1-W2=1.07$~mag and
$W1-W2=1.04$~mag, respectively. The first criterion of Stern \etal
(2012) selects 70 AGNs (71\% of the sample) and the first criterion of
Assef \etal (2012) selects 98 (88\% of the sample). We therefore
adopted the Assef \etal (2012) method to eliminate AGNs from our
sample.

We used the AGN and Galaxy Ages Survey (AGES) deep data (Kochanek \etal
2012) to study the depth of WISE for AGNs in the OGLE-IV Magellanic Bridge
data. We selected objects with AGES flags $QSO=1$ and/or $AGN=1$, spectrum
redshift $z>0.6$ to avoid galaxies, and Spitzer colors
$[3.6]-[4.5]>0.5$~mag to avoid stars and galaxies, as AGNs. The faintest
objects reached $I<22.5$~mag, \ie $\approx2.5$~mag deeper than OGLE. The
Spitzer $[4.5]$ band is nearly identical to WISE $W2$ channel with only 3\%
magnitude differences for AGNs up to $z=4.5$, so we explicitly assumed them
here as equal. The Magellanic Bridge is located close to the South Ecliptic
Pole, where WISE obtained the highest number of epochs. At the center
(further edge) of the Magellanic Bridge, WISE obtained $>30$ ($>20$) images
and the signal-to-noise ratio of 3 is reached at $W1<18.7$~mag and
$W2<17.2$~mag ($W1<18.4$~mag and $W2<17.0$~mag). The average AGN $I-W2$
color is 4.7~mag, with the 1$\sigma$ range of $3.8<I-W2<5.5$~mag. Taking
into account both OGLE and WISE depths, we should be finding 94\% of AGNs,
as compared to fixing the AGES {\it I}-band magnitude to the OGLE limit and
leaving the WISE limit unconstrained.

We matched 1924 galaxies from the OGLE-IV LMC GSEP catalog to the WISE
catalog using a 5\arcs search radius. There were 1658 matches with a median
color $W1-W2=0.22$~mag. We also matched 300 stars with $15<I<17$~mag from
the OGLE-IV LMC GSEP catalog and 228 stars were matched within 3\arcs
radius. Their median color is $W1-W2=-0.12$~mag. In the right panel of
Fig.~8, we show $W1-W2$ histograms for the three types of objects discussed
here. It is clear that even with as simple cut as proposed by Stern \etal
(2012), we can safely remove the majority of AGNs.

\vspace*{-9pt}
\Section{OGLE Transient Detection System}
\vspace*{-4pt}
In October 2012, we implemented the OGLE-IV Transient Detection System
(OTDS) running in a near-real-time at the Warsaw telescope in Las
Campanas Observatory, Chile. Its full description will be presented
in a forthcoming paper, so we only outline the procedure here.

Raw images are reduced on-the fly, but we run the OTDS pipeline to
search for transients just after the last reduction of the night is
done and the databases are updated. The lag between taking an image
and updating the database can then take up to 24 hours. The OTDS works
twofold. First, we inspect a list of new objects with at least two
detections at the same location occurring in subsequent images
daily. Currently, we only investigate {\it I}-band images as the bulk
($\approx90\%$) of observations are obtained with this filter. This
includes checking the light curves and both template, original and
subtracted (difference) images. We produce light curves for all
promising objects to check their behavior prior to the candidate
transient. Second, we will search for SNe in the standard database
(not implemented yet). Both approaches will return real SNe or Novae
but also spurious detections, such as cosmic rays, photometric
problems as well as sometimes variable stars, AGNs and
asteroids. Cosmic rays (hitting twice at the same spot!) and
subtraction artifacts are by far the most common sources of
contamination and are removed automatically using a self-organizing
map (SOM) technique (\eg Wyrzykowski and Belokurov 2008). We then
concentrate on discriminating SNe from other real sources such as AGNs
(Section~4).

During last three months of 2012, we discovered (or co-discovered) 52
SNe using OTDS. Several of them have been spectroscopically observed
and subsequently confirmed by PESSTO collaboration. For example,
OGLE-2012-SN-032 and OGLE-2012-SN-007 turned out to be of Type~Ia
(Anderson \etal 2012b), OGLE-2012-SN-005 is also SN Type~Ia
(Anderson \etal 2012a), and OGLE-2012-SN-032 is either SN Type~Ia or
Ic (Marchi \etal 2012). OGLE-2012-SN-014 (SN Ia) is also known as SN
2012fu (Maza \etal 2012) and OGLE-2012-SN-009 (SN Ia) that peaked at
$I\approx14.9$~mag is known as SN 2012dk (Bock, Parrent, and Howell
2012). OGLE-2012-SN-051 is a Type Ia SN (Taddia \etal 2013) so is
OGLE-2012-SN-049, while OGLE-2012-SN-048 and OGLE-2012-SN-050 are both
SN Type IIn (Sollerman \etal 2013).

\Section{Summary}
In this paper, we presented the results of our search for transients
in the OGLE-IV fields located between the Magellanic Clouds. We
discovered 126 SNe, two dwarf novae, and two AGNs, while inspecting
$\approx70\,000$ pre-selected two-year-long light curves from
$\approx65$~deg$^2$ of the OGLE-IV Magellanic Bridge data. Based on
the known SNe rates, SNe absolute magnitudes, galactic and host
extinctions, and calculated {\it K}-corrections, we simulated the
expected numbers of SNe in our survey. We then compared our cumulative
number of SNe as a function of {\it I}-band magnitude to the simulated
one, finding that our SNe detection efficiency is very high for
$I<18.8$~mag.

Since establishing OTDS in October 2012, we have discovered 52 new
SNe. This number will increase quickly, as the number of observed
fields increases to a total of $\approx600$~deg$^2$. All these
discoveries are published on-line on the OGLE web page. In the future,
for the brightest and spectroscopically confirmed SNe it will be
possible to create high-cadence, well-calibrated light curve templates
for various SNe types.  All data presented in this paper are available
to the astronomical community from the OGLE Internet archive
accessible from the OGLE WWW Page or directly:

\begin{center}
{\it http://ogle.astrouw.edu.pl}\\
{\it ftp://ftp.astrouw.edu.pl/ogle/ogle4/transients/SN/MBR}
\end{center}

Please read the {\sc README} file for the details on the data presented
there as well as on all updates.

The OTDS transients are available from the following webpage:
\begin{center}
{\it http://ogle.astrouw.edu.pl/ogle4/transients/}
\end{center}

\Acknow{We would like to thank Dr.\ Jose L.\ Prieto for many stimulating 
discussions, Prof.\ Christopher S.\ Kochanek and Prof.\ Krzysztof
Z. Stanek for comments on the early draft.

The OGLE project has received funding from the European Research
Council under the European Community's Seventh Framework Programme
(FP7/2007-2013)/ERC grant agreement no.\ 246678 to AU.

The research in this paper was partially supported by the Polish
Ministry of Science and Higher Education through the program ''Ideas
Plus'' award No. IdP2012 000162 to I.S.

This publication makes use of data products from the Wide-field
Infrared Survey Explorer (WISE), which is a joint project of
the University of California, Los Angeles, and the Jet Propulsion
Laboratory/California Institute of Technology, funded by the National
Aeronautics and Space Administration.}

\end{document}